\documentclass[aps, reprint,superscriptaddress]{revtex4-1}
\usepackage{graphicx}%
\usepackage{xcolor}
\usepackage{here}
\usepackage{dcolumn}
\usepackage{bm}
\usepackage{amsmath,amssymb}
\usepackage{color} 
\usepackage{epstopdf}
\usepackage{soul}
\usepackage{interval}
\usepackage[normalem]{ulem}

\definecolor{dred}{rgb}{0.0, 0.5, 0.0}
\definecolor{rcolor}{rgb}{0.9, 0.12, 0.5}

\definecolor{dgreen}{rgb}{0.8, 0.1, 0.02}

\newcommand{\eqn}[1]{
\begin{eqnarray}
	#1
\end{eqnarray}
}
\newcommand{\rc}[1]{\textcolor{black}{{#1}}}
\newcommand{\rrc}[1]{\textcolor{black}{{#1}}}
\newcommand{\tblue}[1]{\textcolor{black}{#1}}

\usepackage{newtxtext,newtxmath}
\usepackage{times}
\usepackage{comment}
\usepackage{hyperref}

\begin{document}

\title{Probing XY phase transitions in a Josephson junction array with tunable frustration}

\author{R.~Cosmic}\email{cosmic@qc.rcast.u-tokyo.ac.jp}
\affiliation{Center for Emergent Matter Science (CEMS), RIKEN, Wako, Saitama 351-0198, Japan}
\affiliation{Research Center for Advanced Science and Technology (RCAST), The University of Tokyo, Meguro-ku, Tokyo 153-8904, Japan}

\author{K.~Kawabata}
\affiliation{Department of Physics, The University of Tokyo, 7-3-1 Hongo, Bunkyo-ku, Tokyo 113-0033, Japan}

\author{Y.~Ashida}
\affiliation{Department of Physics, The University of Tokyo, 7-3-1 Hongo, Bunkyo-ku, Tokyo 113-0033, Japan}
\affiliation{Department of Applied Physics, University of Tokyo, 7-3-1 Hongo, Bunkyo-ku, Tokyo 113-8656, Japan}

\author{H.~Ikegami}\email{hikegami@riken.jp}
\affiliation{Center for Emergent Matter Science (CEMS), RIKEN, Wako, Saitama 351-0198, Japan}

\author{S.~Furukawa}
\affiliation{Department of Physics, The University of Tokyo, 7-3-1 Hongo, Bunkyo-ku, Tokyo 113-0033, Japan}
\affiliation{Department of Physics, Keio University, 3-14-1 Hiyoshi, Kohoku-ku, Yokohama 223-8522, Japan}

\author{P.~Patil}
\affiliation{Department of Physics, Boston University, 590 Commonwealth Avenue, Boston, Massachusetts 02215, USA}

\author{J.~M.~Taylor}
\affiliation{Joint Center for Quantum Information and Computer Science (QuICS), University of Maryland, College Park, Maryland 20742, USA}
\affiliation{Joint Quantum Institute (JQI), National Institute of Standards and Technology, Gaithersburg, Maryland 20899, USA}

\author{Y.~Nakamura}\email{yasunobu@ap.t.u-tokyo.ac.jp}
\affiliation{Center for Emergent Matter Science (CEMS), RIKEN, Wako, Saitama 351-0198, Japan}
\affiliation{Research Center for Advanced Science and Technology (RCAST), The University of Tokyo, Meguro-ku, Tokyo 153-8904, Japan}

\date{\today}

\begin{abstract} 
The seminal theoretical works of Berezinskii, Kosterlitz, and Thouless presented a new paradigm for phase transitions in condensed matter that 
are driven by topological excitations.
These transitions have been extensively studied in the context of 
two-dimensional
XY models -- coupled compasses -- and have generated interest in the context of quantum simulation. Here, we use a circuit quantum-electrodynamics architecture to study the critical behaviour of engineered XY models through their dynamical response. In particular, we examine not only the unfrustrated case but also the fully-frustrated case
which leads to enhanced degeneracy associated with the spin rotational \rc{[U$(1)$]} and discrete chiral \rc{($Z_2$)} symmetries. 
The nature of the transition in the frustrated case has posed a challenge for theoretical studies while direct experimental probes remain elusive. 
Here we identify the transition temperatures for both the unfrustrated and fully-frustrated XY models by probing a Josephson junction array close to equilibrium using weak microwave excitations and measuring the temperature dependence of the effective damping obtained from the complex reflection coefficient. \rc{We argue that our probing technique is primarily sensitive to the dynamics of the U$(1)$ part.}
\end{abstract}

\maketitle

\section{Introduction}
The Berezinskii-Kosterlitz-Thouless (BKT) mechanism provides a prototypical example of topological phase transitions in two-dimensional systems. The transition is accompanied by unbindings of topological defects, known as vortices~[Fig.~\ref{Fig1}(c)]~\cite{berezinskii1972destruction,kosterlitz1973ordering,kosterlitz1974critical}, and have stimulated interest in the context of quantum simulation~\cite{hung2011observation,yamamoto2014exciton,bkt2015nonequilibrium,berloff2017realizing,leo2018collective,king2018observation}. The BKT transition has been extensively investigated in various systems such as thin He films~\cite{bishop1978study}, bosonic and fermionic cold atoms~\cite{hadzibabic2006berezinskii,dalibard2012superfluid,murthy2015observation}, thin film superconductors~\cite{resnick1981kosterlitz,abraham1982resistive,van1987phase,martinoli2000two}, \rc{exciton-polariton} system~\cite{roumpos2012power} and a qubit array~\cite{king2018observation}. One of the representative models for investigating the BKT physics is the two-dimensional XY model---not only the unfrustrated case, where the conventional BKT transition is known to take place, but also cases with frustration in spin interactions have been investigated. In particular, in the fully-frustrated XY case~\cite{teitel1983phase}, the alternating pattern of the chirality [see Fig.~\ref{Fig1}(d)] leads to the enhanced degeneracy of the ground state associated with continuous U$(1)$ and discrete $Z_2$ symmetries, giving rise to potentially two phase transitions at different temperatures corresponding to each symmetry. This frustrated model plays a fundamental role in a variety of frustrated magnetic systems, including spin glasses~\cite{kawamura2010chirality}, clock models~\cite{clockmodel}, and in general, any critical points which possess multiple symmetries. Yet, the exact nature of its phase transition has long remained elusive despite extensive theoretical investigations~\cite{teitel1983phase,Gary@1989,Thijssen1990,Enzo1991,nicolaides1991monte,Jorge1992,Granato1993,Knops1994,Sooyeul1994,olsson1995two,Nightingale1995,luo1998dynamic,hasenbusch2005multicritical,okumura2011spin,huijse2015emergent,lima2018fully}. \rc{On the experimental side, some realizations of the frustrated models have been achieved in ultracold atomic Bose gases in optical lattices~\cite{struck2013engineering,kennedy2015observation}. While enhanced ground-state degeneracy due to $Z_2$ symmetry and its spontaneous breaking have been observed in these systems, 
a detailed nature of the finite-temperature transition is left unaddressed.} In addition, recent progress in numerical method has reignited interest in dynamical features of the quantum XY model~\cite{nandini2014dynamical}.

Although the BKT physics has been observed in various systems, a direct realization of the XY model is limited. One of the attractive platforms to realize the XY model is Josephson junction arrays (JJAs)~\cite{Fazio_PhysRep2001,newrock2000two,resnick1981kosterlitz,abraham1982resistive,van1987phase}. In JJAs, small superconducting islands are connected by Josephson junctions, where the order-parameter phases of the islands are mapped onto the XY spins. Frustration can also be introduced by applying a magnetic field normal to the JJA. In these systems, the phase transitions in the unfrustrated and fully-frustrated cases have long been studied experimentally mostly by DC or low-frequency measurements of current-voltage ($I-V$) characteristics~\cite{resnick1981kosterlitz,Fazio_PhysRep2001,newrock2000two,abraham1982resistive,van1987phase}, demonstrating the universal jump in the exponent of $I-V$ characteristics for the unfrustrated case, as expected for the BKT transition~\cite{resnick1981kosterlitz,abraham1982resistive,newrock2000two}. However, the nature of the transitions has yet to be clarified especially for the fully-frustrated case---even the transition temperature do not agree between the experiment and the theory~\cite{van1987phase,martinoli2000two}. This is possibly because these experiments have detected highly averaged out-of-equilibrium quantities even if a small \rc{bias} current is used. Recent developments in superconducting circuit-QED techniques offer the possibilities of overcoming this limitation since the system is only weakly disturbed by a small microwave excitation, allowing one to probe dynamical response of the system close to equilibrium~\cite{Glazman2015,cosmic2018circuit,jared@2019}. Here we present a circuit-QED-based study of a JJA to observe thermodynamic signatures, and demonstrate its capabilities in identifying the phase transitions in the unfrustrated and fully-frustrated XY models.

\section{Unfrustrated and frustrated XY models in JJA}
\begin{figure}[H]
\begin{center}
\includegraphics[keepaspectratio]{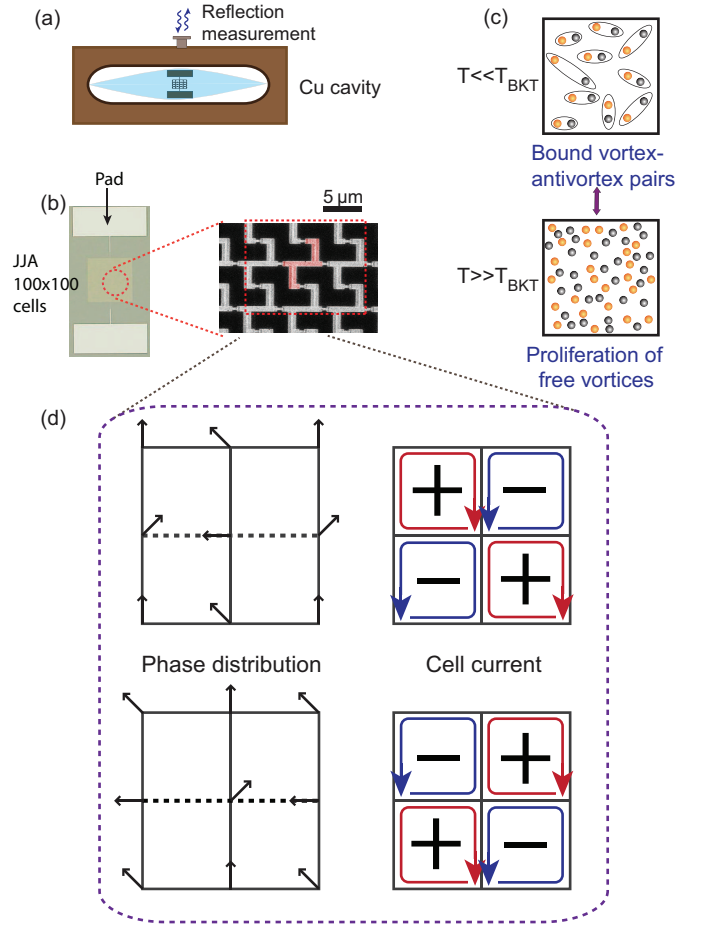}
\end{center}
\caption{\label{Fig1} 
Circuit-QED setup to investigate unfrustrated and frustrated XY models. (a)~Schematic illustration of a JJA using the circuit QED architecture. The JJA is mounted in a 3D microwave cavity. Response of the cavity is measured 
via microwave reflection through the port. (b)~JJA consisting of 100$\times$100 plaquettes connected to two pads. Response of the cavity is investigated via microwave reflection through the port. (c)~\rc{Schematic illustration of the BKT mechanism. Vortex-antivortex pairs are formed at low enough temperatures $(T \ll T_{\rm BKT})$ and all the vortex pairs are unbound to form free vortices at sufficiently high temperatures $(T \gg T_{\rm BKT})$. Note that, near but below $T_{\rm BKT}$, vortex pairs with large distances are excited. At temperatures just above $T_{\rm BKT}$, large-distance pairs dissociate, generating free vortices. We further note that the vortex pairs should overlap in space around $T_{\rm BKT}$.} (d)~Phase distribution (left) and the corresponding current distribution (right) in the degenerate ground states of the fully-frustrated XY model in the Landau gauge~\cite{teitel1983phase}. The arrows in the left panels indicate superconducting phases of the superconducting islands. The dotted horizontal segments represent the antiferromagnetic bonds. The two states are characterized by the opposite senses of chirality shown in the right panel, such as (\textbf{$+$}) and (\textbf{$-$}). }
\end{figure}

\rc{The JJA can be mapped onto the XY model as follows. Each island $i$ [zoomed up and false-colored in Fig.~\ref{Fig1}(b)] is characterized by the phase $\phi_{i}$ of the order parameter and the number of Cooper pairs $n_i$. The Hamiltonian of the JJA~\cite{Fazio_PhysRep2001} is then given by
\begin{equation}\label{eq:hamiltonian}
H_{\rm JJA} = \frac{(2e)^2}{2} \sum\limits_{\left\langle {i,j} \right\rangle} n_i C^{-1}_{ij} n_j - E_J\sum\limits_{\left\langle {i,j} \right\rangle} \cos(\phi_i-\phi_j-A_{ij}) .
\end{equation}
The first term in Eq.~(\ref{eq:hamiltonian}) represents the charging energy and the second term describes the Josephson effect characterized by the Josephson energy $E_J$, where $C_{ij}$ is an element of a capacitance matrix $\bf{C}$; $ A_{ij}  = \left( {2\pi/\Phi_0 } \right)\int_i^j {\vec{A}} \cdot d{\vec{l}} $ is the line integral of the vector potential $\vec{A}$ from an island $i$ to an island $j$ with $\Phi_{0}=h/2e$ being flux quantum ($e$ is the elementary charge and $h$ is Planck's constant). The main contribution to the charging energy comes from the capacitance between two neighbouring islands $C_J$, which is given by $E_C=e^2/2C_J$.
In the case of $E_J/E_C \approx 2.2 \gg 2/\pi^2$~\cite{Fazio_PRB1991, newrock2000two, Fazio_PhysRep2001} as in our JJA, the charging term is insignificant and can be neglected when equilibrium properties are concerned (however, this term is taken into account when discussing dynamical properties later). 
Then, an isomorphic mapping is possible from the local phases $\phi_i$ onto two-component planar spin variables $S_i=\left(\cos{\phi_i}, \sin{\phi_i}\right)$ at site $i$, where the spin-spin coupling is characterized by $E_J$ and $\phi_{i} \in \mathclose[0,2\pi \mathopen)$. The effective Hamiltonian for zero flux is then given by
\begin{equation}\label{eq:mappedhamiltonian}
H_{\rm{eff}}=-\sum_{\langle i,j\rangle} J_{ij}\vec{S_{i}}\cdot\vec{S_{j}}.    
\end{equation}
Here the interaction is $J_{ij}= J$ [$(=E_{J})>0$ in our notation] for the unfrustrated case. In a magnetic field, the vector potential term gives a frustration for spin configuration. The frustration is maximum for a half flux. For this fully frustrated case, $J_{ij}$ in Eq.~(\ref{eq:mappedhamiltonian}) can have two values $J_{ij}=\pm J$ such that $\prod_\Box J_{ij}/J=-1$~\cite{okumura2011spin}, where the product is taken over a plaquette. In the latter case, one of four bonds in each plaquette is antiferromagnetic as indicated in Fig.~\ref{Fig1}(d), which makes spin configurations frustrated. Then the ground state is not unique but doubly degenerate (apart from the trivial degeneracy under the global U$(1)$ rotation) as illustrated in Fig.~\ref{Fig1}(d).}

\begin{figure*}
\begin{center}
\includegraphics[keepaspectratio]{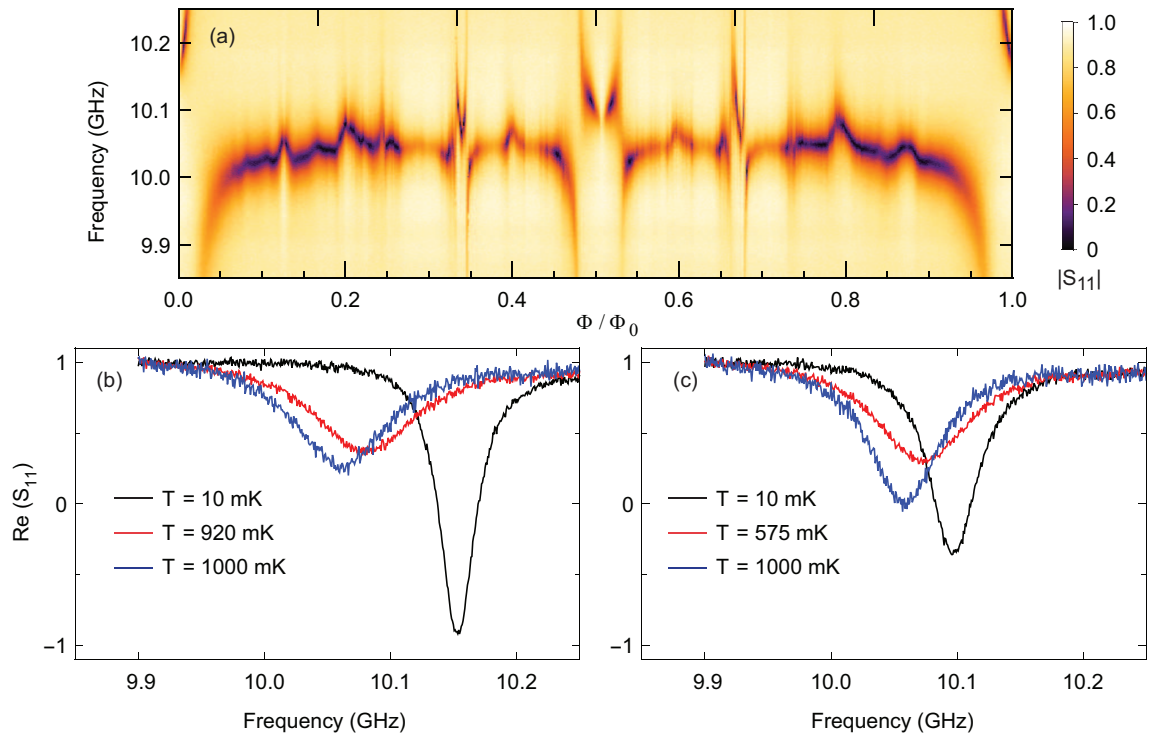}
\end{center}
\caption{\label{Fig2} Reflection spectra of the cavity containing the Josephson junction array. (a)~Absolute value of the reflection coefficient, $|S_{11}|$, as a function of the magnetic-flux bias $\Phi/\Phi_{0}$ and the frequency at 10 mK. The bare cavity frequency is located
around $10.056$ GHz. (b) and (c)~Real part of $S_{11}$ as a function of the frequency for $\Phi/\Phi_{0}=0$ and $1/2$, respectively. Data are taken at a power of $P_{\rm MW}= -132$ dBm ($6.3\times 10^{-17}$~W) at the input port of the cavity.}
\end{figure*}

\section{Experimental methods}
We use a large square network of Josephson junctions made of Al films evaporated on a silicon substrate with $100\times100$ plaquettes shunted by large capacitance electrodes [Fig.~\ref{Fig1}(b)]. The area enclosed by one plaquette is $6\times6\;\mu $m$^2$. The Josephson energy of a single junction is $E_J/h=30.3$ GHz estimated from the resistance at 4 K~\cite{Ambegaokar_PRL1963}, and the charging energy is $E_c/h=13.8$~GHz. The shunting capacitance $C_S=48.5$ fF from the antenna electrodes [Fig.~\ref{Fig1}(b)] which also provide coupling to the cavity in the same manner as in a typical transmon qubit~\cite{Koch_PRB2007}. \rc{The JJA has two different kinds of edges: At two of the four edges which are connected to a pad, islands are galvanically connected without Josephson junctions, while islands at the other two edges are connected with Josephson junction as same to other part. Note that probing of the JJA by microwave is done through the galvanically-connected edges.} The estimated ground capacitance of an island is $C_g=1.38$ aF. The full capacitance matrix is extracted using finite element analysis, and the junction capacitance is estimated from the capacitance density, i.e., capacitance per area of $60$ fF$/ \mu$m$^2$. The transition temperature of the superconducting film is determined to be $1.375$~K by a low-frequency four-probe measurement on one of the antenna pads. The temperature dependence of $E_J (T)$ is determined using the standard Ambegaokar-Baratoff relation~\cite{Ambegaokar_PRL1963} combined with the temperature dependence of the gap which is numerically found using the BCS relation by setting the transition temperature of the Al film. The JJA is placed in the center of a 3D microwave cavity made out of OFHC copper [Fig.~\ref{Fig1}(a)], where the electric-field strength of the fundamental mode (TE$_{101}$) is the strongest. The bare resonance frequency is at $10.056$~GHz. The coupling between the cavity and the JJA~\cite{bourassa2012josephson} is estimated to be $g/2\pi=350$~MHz.  

To investigate the dynamical response of the JJA to microwaves, we measure the reflection spectrum of the cavity under weak driving with the external coupling rate of $\kappa_{\rm ext}/2\pi=31.8$~MHz. We study the dynamical response of the XY models at a single-photon level by monitoring the complex reflection coefficient in the microwave spectroscopy through the cavity (Appendix~\hyperref[measurement_setup]{A}). This acts as a probe of vortex structures~\cite{cosmic2018circuit} and the critical temperature as it is sensitive to vortex excitations. \rc{We note that the dipole interaction between the cavity and the JJA can be treated perturbatively when the photon number inside the cavity is limited to the  single-photon level~(Appendices~\hyperref[theory]{B} and \hyperref[input_output]{C}).}

\section{Concept of the experiment}
First, we describe a theoretical idea of what information can be obtained by the circuit-QED setup. Since $E_J/E_C\gg 2/\pi^2$, it is useful to describe the dynamics of the JJA using Hamilton's classical equations of motion in terms of the phases $\phi_i$ and the charges $q_i=2en_i$ (conjugate momentum of $\hbar\phi_i/(2e)$). We further include thermal Langevin noises (with the damping rate $\Gamma$) in the equations to model possible dissipation processes in the experimental system. With such noises, the finite-temperature Gibbs distribution is obtained after the convergence to a steady state. 
In this formulation, the susceptibility relating the perturbation by external microwaves to the charge $q_0$ at the top pad is given by~(Appendix~\hyperref[input_output]{C})
\begin{equation}\label{eq:susceptibility}
\chi(\omega,T)=\frac{\hbar g}{{2}e}\int_{-\infty}^{+\infty}dt \, \, e^{i\omega t}\frac{\langle {q}_0(t)\dot{{q}}_0(0)\rangle_\mathrm{eq}}{k_B T}.
\end{equation}
Here, 
$g$ characterizes the coupling strength to the microwave 
and $\langle \cdots\rangle_\mathrm{eq}$ represents the thermal average. 
We note that dissipation processes manifest themselves in the imaginary part of $\chi(\omega,T)$. The complex reflection coefficient $S_{11}$ is related to the susceptibility using the input-output formalism~\cite{cosmic2018circuit} (Appendix~\hyperref[input_output]{C}): 
\begin{equation}\label{eq:input-output}
S_{11} =-\frac{
\frac12\left(\kappa_{\rm ext}-\kappa_{\rm int}-g\frac{ \rm{Im}\chi(\omega)}{2e}\right)+i \left(\omega-\omega_{c}+\frac{g}{2}\frac{\rm{Re}\chi(\omega)}{2e}\right)}{\frac12\left(\kappa_{\rm ext}+\kappa_{\rm int}+g\frac{ \rm{Im}\chi(\omega)}{2e}\right)-i \left(\omega-\omega_{c}+\frac{g}{2}\frac{\rm{Re}\chi(\omega)}{2e}\right)}.
\end{equation}
We can define a damping coefficient (``linewidth'') for the cavity using $S_{11}$ and obtain $\kappa_{\rm tot} = \kappa_{\rm ext}+\kappa_{\rm int}+g{\rm Im}\chi(\omega_c)/2e$, where $\kappa_{\rm int}$ is the internal loss rate, and $\kappa_{\rm ext}$ is the external loss rate due to the coupling of the cavity to the input port. The key quantity of interest is $ {\rm Im}\chi(\omega_c)/2e$, which can be extracted from experimentally obtained $\kappa_{\rm tot}$. \rc{As we will show later, ${\rm Im}\chi$ is sensitive to the U$(1)$ part of the transition.} We note that $\kappa_{\rm ext}$ and $\kappa_{\rm int}$ do not depend on $\Phi$ or $T$ in our setup, as we verified it by a measurement with a plain silicon substrate put in the cavity and simultaneously fitting the real and imaginary part of $S_{11}$. The change in $\kappa_{\rm tot}$ as a function of $T$ represents the change in the internal loss of the JJA, or more precisely, it reflects the dependence of ${\rm Im}\chi(\omega_c)$ on $T$.

\section{Probing dynamical Susceptibility}
We measure the cavity reflection while applying a magnetic field perpendicular to the plane of the array in order to study the frustration-induced properties of the system~(Appendix~\hyperref[measurement_setup]{A}). As shown in Fig.~\ref{Fig2}(a), the spectrum is rich and has various features at commensurate flux values such as $\Phi/\Phi_0=0$, $1/3$, and $1/2$, and exhibits reflection symmetry about the half flux case. The flux-dependent dispersive shift of the line center from the bare resonance of the cavity at $10.056$~GHz is observed due to the non-topological collective oscillations of $\vec{\phi}$ around the lowest-energy configurations of phases $\vec \phi_{0}$---plasma modes. Here $\vec \phi = (\phi_1,~\cdots ,~ \phi_n)^{\rm T}$. This pattern of the shift indicates that vortices induced by the magnetic field form a rigid lattice-ordered state around the commensurate flux value~\cite{cosmic2018circuit}. Below we focus on the results at $\Phi /\Phi_0=$ 0 and 1/2, and discuss how the spectrum evolves when temperature is increased.

\begin{figure}
\begin{center}
\includegraphics[keepaspectratio]{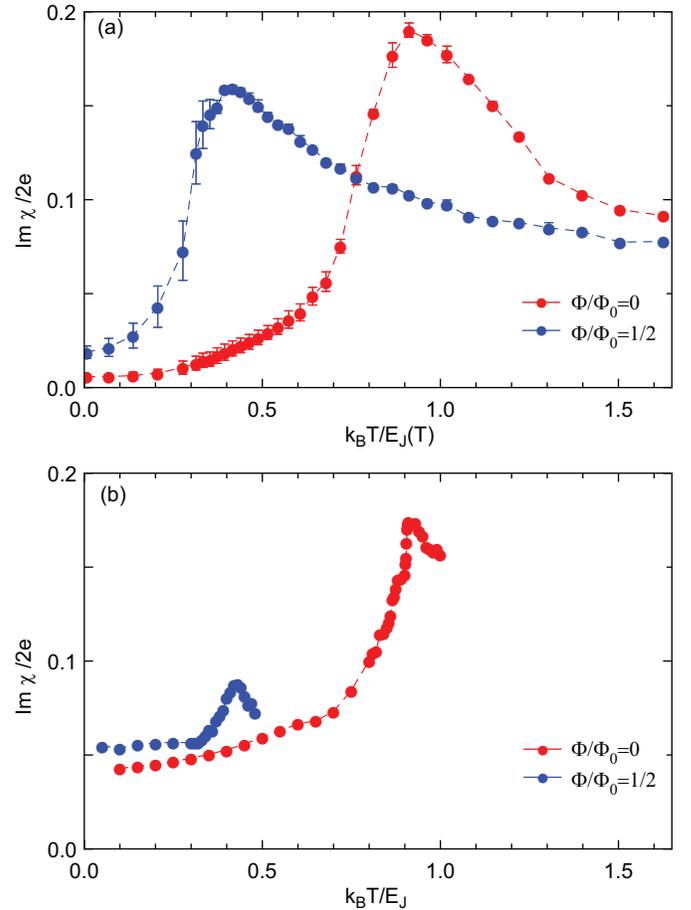}
\end{center}
\caption{\label{Fig3} Dynamical susceptibility. (a)~${\rm Im}\chi$ as a function of $T/E_{J}(T)$ obtained in the experiment of a $100\times 100$ JJA at $\Phi/\Phi_{0}=0$ and $1/2$. (b)~Simulated susceptibility  ${\rm Im}\chi$ at the cavity frequency $\omega_{\rm c}$ as a function of the reduced temperature $k_{\mathrm{B}} T/E_{\mathrm{J}}$. A lattice of $50\times50$ plaquettes connected to two pads is used in this simulation.
}
\end{figure}

Figures~\ref{Fig2}(b) and (c) show the real part of the reflection spectra for $\Phi/\Phi_{0}=0$ and $1/2$ for three different temperatures, one below, one close to, and one above the transition temperature defined below. The linewidth is narrow at 10 mK, but becomes broader with increasing temperature. Further increasing temperature, the linewidth becomes narrower again. This change of the linewidth originates from the change in $\rm{Im}\chi$. The extracted $\rm{Im}\chi$ by fitting the spectrum to Eq.~(\ref{eq:input-output}) is shown in Fig.~\ref{Fig3}(a) as a function of the normalized temperature $k_{B}T/E_J(T)$. The linewidth is found to exhibit a peak at a certain $k_{B}T/E_J(T)$ for both $\Phi/\Phi_{0} =$ 0 and 1/2. \rc{[We note that $E_{J}(T)$ is suppressed at high temperatures, but even at the peak temperature of the zero-flux case, $E_{J}(T)/E_{C}=1.52$ $\gg 2/\pi^2$, indicating the JJA is well in classical regime.]} This indicates that the linewidth of the cavity is a key quantity to characterize the system upon heating.

\begin{figure}
\begin{center}
\includegraphics[keepaspectratio]{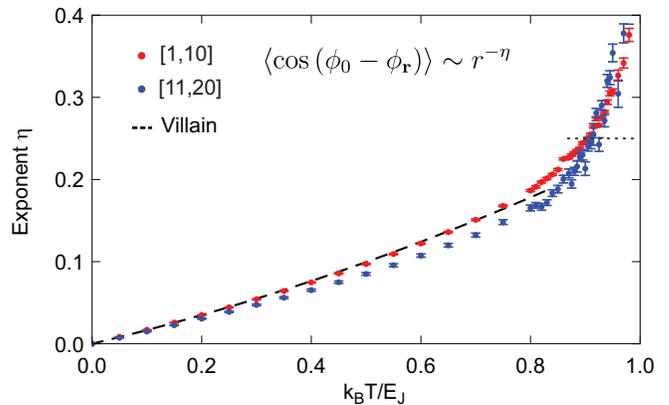}
\end{center}
\caption{\label{Fig4} 
Critical exponent $\eta(T)$. Numerically simulated critical exponent as a function of the reduced temperature $k_BT/E_J$. The exponent is extracted from linear fitting of the log-log plot of the spatial correlation function in the ranges of the lattice intervals $[1,10]$ (red) and $[11,20]$ (blue) (Appendix~\hyperref[theory]{B}). The black dashed line shows the perturbative result by Villain~\cite{villain1975theory}. A system size of $100\times100$ and periodic boundary conditions are used in this simulation.}
\end{figure}

These experimental results are compared with the results of our Langevin dynamics simulations of ${\rm Im}\chi(\omega_{c})$ in Fig.~\ref{Fig3}. Here, the stochastic equations of motion are numerically solved, and physical quantities are calculated after the convergence to a steady state. 
The simulated ${\rm Im}\chi(\omega_{c})$ also exhibits a peak for both $\Phi /\Phi_{0}=0$ and $1/2$ as shown in Fig.~\ref{Fig3}(b). \rc{While the calculations are limited to a system size $30\leq L\leq 50$ due to the numerical cost of dealing with the large nondiagonal capacitance matrix, we confirm that the size dependence of the peak temperatures is below $\sim 5$\% (Appendix~\hyperref[theory]{B}).} 
Remarkably, the peak temperatures agree well between the experiments and the simulations for both $\Phi/\Phi_0 = 0$ and $1/2$. For $\Phi/\Phi_0 = 0$, the peak temperature ($k_\mathrm{B}T / E_\mathrm{J}\simeq 0.91$) precisely corresponds to the BKT transition temperature, which is confirmed by the simulated results plotted in Fig.~\ref{Fig4}. At this temperature, the critical exponent $\eta$ for the spin correlation reaches $1/4$, the value expected from the BKT mechanism~\cite{villain1975theory,nelson1977universal,minnhagen1987two}. 
We note that we benchmarked the numerical analysis by confirming the expected critical decay of the correlation function $C(r)=\langle\cos{\left(\phi_{0}-\phi_{\bf r}\right)\rangle}$, with ${\bf r}=(0,r)$ and $r$ is the site distance, in the equilibrium regime for systems with different sizes by including only a diagonal capacitance matrix (see Fig.~\ref{Fig4} and Appendix~\hyperref[theory]{B}). 
For $\Phi/\Phi_0 = 1/2$, the peak of $\rm{Im}\chi$ is found at $k_\mathrm{B}T / E_\mathrm{J}\simeq 0.43$, which is also consistent with the transition temperature identified in previous numerical calculations~\cite{teitel1983phase,Jorge1992,olsson1995two,luo1998dynamic,korshunov2002kink,okumura2011spin}. \rc{We note that the U$(1)$ and $Z_2$ transition temperatures obtained in previous numerical studies, $k_\mathrm{B}T / E_\mathrm{J}\simeq 0.44$~\cite{teitel1983phase,okumura2011spin,lima2018fully}, are close to each other.}
In attempting to compare theoretical and experimental susceptibilities, we found that qualitative agreement was achieved by taking the heuristic damping coefficient $\Gamma$ to be comparable to the kinetic fluctuation scale of the system, $E_C/\hbar$ [specifically, we set $\Gamma=4E_C/\hbar$ in Fig.\ \ref{Fig3}(b)]. Some differences such as the peak height for half flux could be because of additional dissipation processes not captured in the Langevin dynamics approach, particularly with respect to vortex dynamics. We further note that it is challenging for the Langevin dynamics simulation to reproduce the suppression of ${\rm Im}\chi$ at low temperatures as observed in experiment---as the temperature is lowered, an increasingly larger number of discrete time steps is required to achieve sufficient convergence because of the slower decay of the temporal correlation.

\section{Discussion}
The agreement between the numerical and experimental results~(Fig.~\ref{Fig3}) indicates that the observed peak in the linewidth is caused by the phase transitions. The enhancement of the dissipation observed around the transition temperature is understood as follows. As temperature 
\rc{goes beyond} the transition point from the lower side, \rc{vortex pairs with long distances first unbind at temperatures slightly above the transition temperature.} An external perturbation by cavity photons then drives the dynamics of the vortices that subsequently equilibrate by accompanying energy dissipation. It is this interplay between the external drive and the diffusive dynamics of vortices that manifests itself as the pronounced dissipation near the transition temperature (cf.~Fig.~\ref{Fig3}). Further evidence supporting damping via driving of unbound vortices is provided by the increase in linewidth seen even at zero temperature as the flux is moved slightly away from 0 or 0.5, where free vortices induced by the magnetic field cause damping, as is evident from the broad spectrum seen in Fig.~\ref{Fig2}(a). \rc{Upon further heating, vortices pairs with shorter distances separate further and finally unbind. When an oscillating current is induced by the microwave drive, the unbound vortices and antivortices diffusively move to the opposite directions normal to the current, resulting in the screening of the oscillating current. At higher temperatures, more free vortices appear and screen the current more, thus the
vortices become less susceptible to the external perturbation. The process results in
smaller dissipation at higher temperatures.}

The extracted dynamical susceptibility observed in $\rm{Im}\chi$ in Fig.~\ref{Fig3} is a measure of the spectral dynamical correlation
function near equilibrium in the linear response regime. It is known 
that ${\rm Im}\chi$ holds physical information about the system
close to $T_{c}$~\cite{bishop1978study,ambegaokar1978dissipation,glauber1963time}. Below $T_c$, ${\rm Im}\chi$ is sensitive to the
typical size of a vortex bound pair, $\xi_-$, and above $T_c$, its
behavior is controlled by $\xi_+$, the typical distance
between free vortices. This underscores the importance of the
information contained in ${\rm Im}\chi$, as we can extract physically
relevant features of vortices. The BKT physics has been explored for the
low frequency limit for JJAs, and our linewidth measurements open a window
into the frequency-domain response.

\rc{As demonstrated for the zero-flux case (Fig.~3), our experimental scheme is largely sensitive to the U$(1)$ symmetry. This is explained as follows: In our setup, we extract the response functions via the charge $q_0(t)$ induced on one of the pads connected to the JJA. As the response function Eq.~(\ref{eq:susceptibility}) is expressed in terms of $q_{0}(t)$, we expect that the susceptibility is sensitive  to the U(1) phase since the charge $q_{0}(t)$ is proportional to the time derivative of the phase. Because $\phi_0$ is the conjugate variable of $q_0$, the latter can be viewed as a "velocity" of the spin. This is demonstrated for the unfrustrated case in our experiment (cf.~Fig~\ref{Fig3}). Furthermore, a similar detection of the U(1) phase transition has been already demonstrated in the work of superfluid helium using torsional pendulum~\cite{bishop1978study, ambegaokar1978dissipation, vinay1980}.}

\rrc{Although this scheme is sensitive to the U(1) part, it is not straightforward to study the critical behavior near the BKT transition using the present scheme with a fixed probing frequency. However, it could be addressed if we use a broader-band detection scheme such as a bad cavity limit or by replacing the cavity with a waveguide. These modifications allow for probing the response function in a wide frequency range, and we can thus address universal features of the transition that emerge in the low-frequency limit.}

\rrc{We further note that the approach we take is sensitive primarily to the spin-waves in the system~\cite{cosmic2018circuit}, rather than directly coupled to the vortices. Instead, the modification of the effective impedances as the vortices move causes changes in the observed narrow-band response at the cavity frequency. (In our prior publication~\cite{cosmic2018circuit}, we showed theoretically and experimentally that this connection allowed observation of vortex lattices at a variety of fillings factors and inter-vortex distances). Thus we rely upon the indirect measurement of the vortex behavior via their effect on the spin-wave spectrum. This indicates that our experiment probes response of the whole array, that is, there is no particular length scale associated with the probe frequency since the spin-waves are defined for the whole junction array. We also note that even though the spin-waves are not exactly defined at high temperatures, our experiment probes an ensemble average of vortices through broad spin dynamics.}

For the fully-frustrated case, there could be two different transitions: one associated with the onset of the chirality long-range order (discrete Z$_2$ part) and the other with the onset of the spin quasi-long-range order [continuous U(1) part]. Numerical studies have indicated that these indeed occur at close but different temperatures~\cite{teitel1983phase,Jorge1992,olsson1995two,luo1998dynamic,korshunov2002kink,Ito2003,okumura2011spin} while there have also been proposals supporting the onsets of the two orders at the same temperature~\cite{Granato1993,Knops1994,hasenbusch2005multicritical,lima2018fully}. Although the precise nature of the transition is still under debate, we expect that the peak of the dynamical susceptibility (Fig.~\ref{Fig3}) is associated with the onset of the spin quasi-long-range order in the U(1) part as explained above.
\tblue{Indeed, in the Monte Carlo study of Ref.~\citenum{Ito2003}, the spin relaxation dynamics data was used to locate the BKT transition point in the U(1) part while this data showed no sign of a transition at the presumed chirality transition point determined by other methods.}
Theoretically, the transition in the chirality part can be detected as an anomaly in the specific heat, while this quantity only shows smooth behavior at the spin transition \cite{teitel1983phase,okumura2011spin}. \rc{We further note that our scheme is unlikely to detect the globally correlated flip of all the plaquettes at low temperatures, either.}

\rc{One of the main difficulties of FFXY in JJA is that both U(1) and Z$_2$ are defined in terms of the phases of the islands, and thus they could be coupled. Even though Z$_2$ is uncoupled from U(1) due to its discrete symmetry at low temperatures, we do not exclude the possibility that the susceptibility can detect Z$_2$ part close to the transition temperature. However, in view of the argument presented above, we consider that the significant contribution arises from the U(1) part.}  \rc{Finding a suitable quantity for detecting the chirality transition in the circuit-QED setup and investigating the highly nontrivial and debated nature of the transitions experimentally are interesting issues for future studies.}

\section{Conclusion}
In conclusion, we successfully identified the U$(1)$ transition temperatures of the unfrustrated and fully frustrated XY models by studying the dynamical susceptibility in the circuit QED experiment. The experimental system reported here
serves as an ideal platform to study frustrated spin systems in a highly controllable manner. As such, our work also provides a benchmark test for probing the many-body physics, paving the way toward studying challenging regimes of JJAs, in particular in frustrated quantum regime.

\section*{Acknowledgements}

We acknowledge J.M.~Kosterlitz, N.~Nagaosa, T.~Yamamoto, V.~Sudhir, M.~Oshikawa, C.~J.~Lobb, N.~T.~Phuc, Z.R.~Lin, K.~Inomata and T.~Hanaguri for fruitful discussions. We also thank R.S.~Deacon for the experimental support and discussion. This work was partly supported by ImPACT Program of Council for Science, Technology and Innovation and the Matsuo Foundation. R.C. was supported by MEXT. K.K. and Y.A. were supported by the JSPS through Program for Leading Graduate Schools (ALPS).

\section*{Appendix A: Experimental Setup}
\label{measurement_setup}
\begin{figure}
\begin{center}
\includegraphics[keepaspectratio]{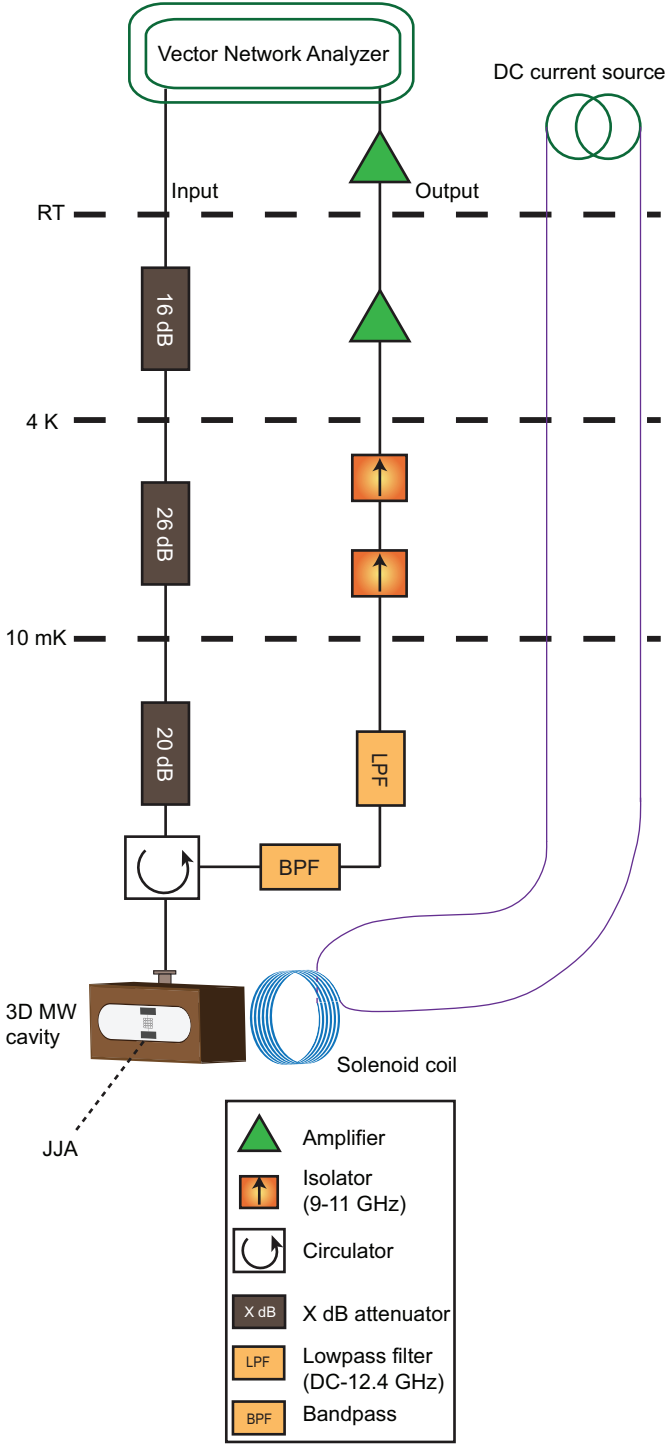}
\end{center}
\caption{\label{FigS1} Schematic diagram of the measurement setup. }
\end{figure}

A schematic of the experimental setup is shown in Fig.~\ref{FigS1}. We measure the complex reflection coefficient of the cavity, $S_{11}$, as a function of frequency using a vector network analyzer (VNA). An input microwave tone is sent to the coaxial cable, which is subsequently attenuated in order to minimize the background noise. The reflected microwave tone from the cavity is amplified by a cryogenic high-electron-mobility transistor (HEMT) amplifier followed by a room temperature amplifier with a total gain of$\sim 66$ dB, which is detected by the VNA. Microwave power presented in the paper refers to that at the cavity port. 

The Josephson junction array (JJA) used in this study consists of $100\times100$ plaquettes and is lithographically fabricated on a silicon substrate by a double angle evaporation of aluminium films. The JJA is connected to two large electrodes in order to provide strong coupling to the $\rm{TE}_{101}$ mode of the 3D cavity. The cavity has higher modes; the next four higher modes are the TE$_{201}$, TE$_{102}$, TE$_{202}$, and TE$_{301}$ modes located at 16.4, 16.7, 20.9, and 21.3 GHz, respectively.
Among these modes, only the TE$_{101}$ and TE$_{301}$ modes dominantly couple to the JJA. The parameters of the system are shown in Table~I. We note that the superconducting-normal transition temperature shown in Table~\ref{tab:1} is obtained by a low-frequency four-probe measurement on one of the shunting pads.

\begin{table}[thb]
\caption{\label{tab:1} System parameters.}
\label{table1}
\begin{tabular}{ccc}
\hline
\hline
Bare cavity resonance frequency & $\omega_c/2\pi$& $10.056\;\rm{GHz}$   \\
\hline
Cavity external coupling rate & $\kappa_{\rm{ext}}/2\pi$& $31.8\;\rm{MHz}$ \\
\hline
Cavity internal loss rate & $\kappa_{\rm{int}}/2\pi$& $0.7\;\rm{MHz}$ \\
\hline
Coupling strength & $g/2\pi$& $350\;\rm{MHz}$ \\
\hline
Area enclosed by one plaquette & $S$& $6\times6\;\mu\rm{m^2}$ \\
\hline
Josephson junction energy & $E_J/h$ & $30.3\;\rm{GHz}$ \\
\hline
Charging energy & $E_C/h$ & $13.8\;\rm{GHz}$ \\
\hline
Shunting capacitance & $C_{S}$ & $48.5\;\rm{fF}$ \\
\hline
Ground capacitance of each island & $C_{g}$ & $1.38\;\rm{aF}$ \\
\hline
Superconducting transition temperature & $T_c$ & $1.375\;\rm{K}$ \\
\hline
\end{tabular}
\end{table}

\setcounter{equation}{0}
\section*{Appendix B: Langevin dynamics simulations}
\label{theory}
\renewcommand{\theequation}{B\arabic{equation}}

\begin{figure}
\begin{center}
\includegraphics[keepaspectratio]{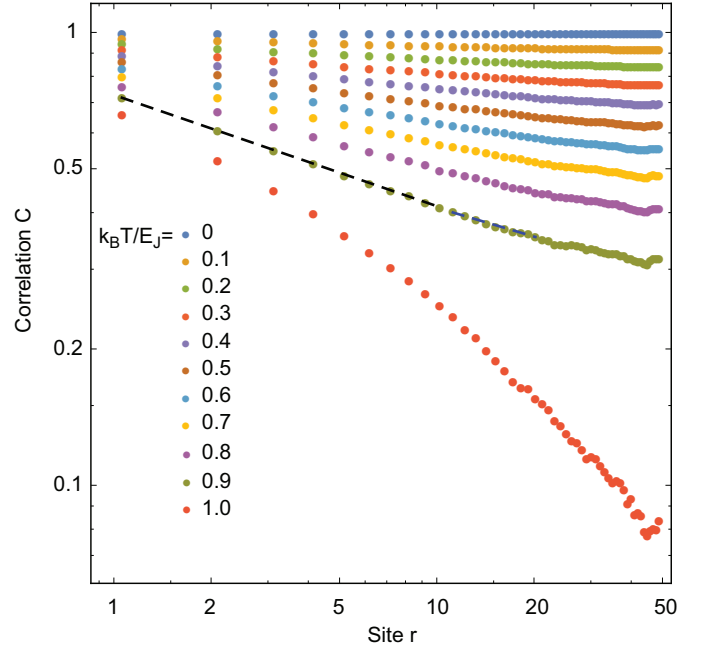}
\end{center}
\caption{\label{FigS2} Critical decay of the correlation function of the XY model. The spatial correlation function $C(r) = \langle \cos \left( \phi_{0} - \phi_{\bf r} \right) \rangle$ in the horizontal direction exhibits power-law behavior as a function of the site distance $r$ below the BKT transition temperature $k_{B} T/E_{\rm J} \sim 0.91$. The numerical calculations are performed for a $100 \times 100$ toy model that satisfies periodic boundary conditions,   
$C_{ij}=C \delta_{ij}$ with $(2e)^2/C=2E_J$, and $\Gamma=E_J$
(we also confirmed that the critical exponents do not significantly depend on $\Gamma$). Decay exponents in Fig.~4 are extracted from linear fitting of the log-log plot of the spatial correlation in the lattice interval of [1,10] (black) and [11,20] (blue) as shown for $k_{B} T/E_{J} = 0.91$.}
\end{figure}
We here describe the details about the Langevin dynamics simulations. To analyze the JJA in the large-$E_J$ limit (i.e., $E_J/E_C\gg 2/\pi^2$), we start from the classical Hamiltonian
\begin{equation}
\begin{split}
H_{\rm JJA} =&\frac{1}{2}\sum_{ij}{q}_{i}C_{ij}^{-1}{q}_{j}-E_{J}\sum_{i,\mu}\cos\left(\phi_{i}-\phi_{i+\mu}+A_{i,\mu}\right)\\ &+ V(t),
\end{split}
\end{equation}
where $\phi_i$ is the phase variable at the lattice site $i$ and $q_i$ is the conjugate momentum (the charge) of $\phi_i/\eta$ with $\eta=2e/\hbar$. The first term is the charging energy with $C_{ij}$ being the capacitance matrix. The second term is the potential energy for the phase variables with $A_{i,\mu}$ being the integral of the vector potential along the segment in the direction of $\mu$ from the site $i$. The last term $V(t)$ represents the time-dependent small perturbation by an external microwave and can be described as 
\eqn{
V(t)=-E(t)\ell q_0,~~E(t)\ell=\frac{\hbar g}{2e}X(t), 
}
where $q_0$ is the charge at the top pad, $E(t)\ell$ is the electric potential difference between the two pads, $g$ is the coupling strength, and $X(t)=(a+a^\dagger)/\sqrt{2}$ is the quadrature of the photon field $a$ (see below). 

To simulate the finite-temperature dynamics of the experimental system, we numerically solve the following stochastic differential equations of motions with the thermal Langevin noises \footnote{We here use the facts that the capacitance matrix is real-symmetric and thus can be diagonalized via an orthonormal matrix and that a linear superposition of the Wiener processes is also the Wiener process.}:
\begin{equation}
\begin{split}
d\phi_{i}&=\eta\sum_{j}C_{ij}^{-1}{q}_{j}dt,\\
\frac{1}{\eta}d{q}_{i}
=&-E_{J}\sum_{\mu}\sin(\phi_{i}-\phi_{i+\mu}+A_{i,\mu})\,dt\\
&+
E_{J}\sum_{\mu}\sin(\phi_{i-\mu}-\phi_{i}+A_{i-\mu,\mu})\,dt\\
&-\frac{\Gamma} {\eta}q_{i} dt+\sqrt{2\Gamma k_{B}TI}\,dW_{i},
\end{split}
\end{equation}
where $\Gamma$ is the damping rate, $dW_i$ is the Wiener stochastic process obeying $\langle dW_{i}dW_{j}\rangle=\delta_{ij}dt$, $T$ is the temperature of the system, and $I=C/\eta^2$ is moment of inertia that does not affect the steady state and can be taken to be an arbitrary value~\cite{LR87}. The steady-state distribution is given by the following Gibbs distribution \cite{LR87}
\eqn{
P(\{\phi_{i}\})\propto\exp\left(\frac{E_{J}}{k_{B}T}\sum_{i,\mu}\cos\left(\phi_{i}-\phi_{i+\mu}+A_{i,\mu}\right)\right),
    \label{eq: def-Gibbs}
} 
where we have integrated out the momentum degrees of freedom. We have benchmarked the convergence to the equilibrium distribution by confirming the expected critical decay of the correlation function. Figure~\ref{FigS2} shows the numerically obtained spatial correlation function
\begin{equation}
C \left( r \right)
= \langle \cos \left( \phi_{0} - \phi_{\bf r} \right) \rangle,~~
{\bf r}=(0,r)
    \label{eq: def-corr}
\end{equation}
as a function of the distance $r$ for a toy model with a diagonal capacitance matrix and periodic boundary conditions. Below the BKT transition temperature $k_{B} T/E_J \sim 0.91$, the correlation function indeed decays according to a power law and its critical exponent obeys the result by the renormalization group~\cite{villain1975theory} as shown in Fig.~4 in the main text. We note that the steady-state distribution $P \left(\{ \phi_{i} \} \right)$ in Eq.~(\ref{eq: def-Gibbs}) and the spatial correlation function $C \left( r \right)$ in Eq.~(\ref{eq: def-corr}) do not depend on the damping rate $\Gamma$ as long as it is nonzero and stochastic realizations are sufficiently many.
\begin{figure}
\begin{center}
\includegraphics[keepaspectratio]{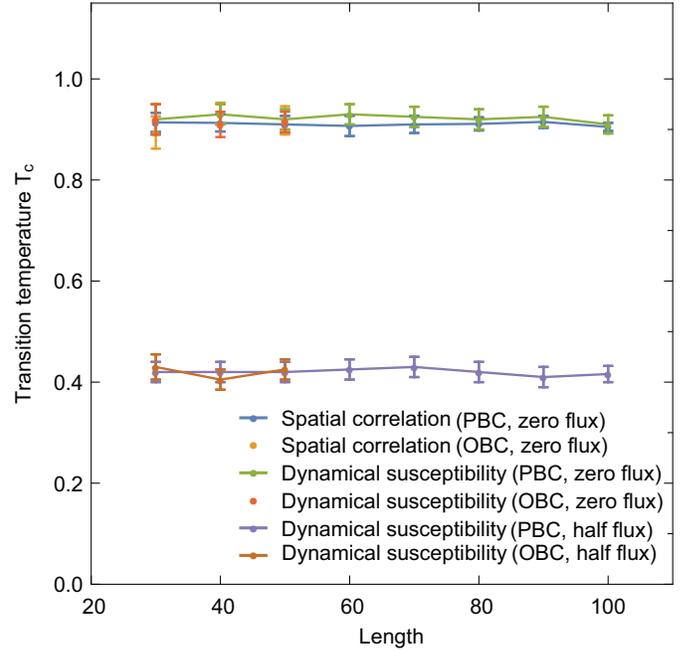}
\end{center}
\caption{\label{FigS3} Size $L$ dependence of the BKT transition temperatures and the peak temperatures of the linewidth. Here OBC (PBC) stands for open (periodic) boundary conditions. The BKT transition temperatures are determined by the critical exponent of the spatial correlation function and shown for the realistic model with open boundaries up to $L=50$ and the toy model with periodic boundaries up to $L=100$ without fluxes. The peak temperatures of the linewidth are plotted.}
\end{figure}

Having established the convergence to the steady-state thermal distribution, we now consider obtaining the linear response function corresponding to an external perturbation by microwaves. Since the dominant contribution in the cavity--array interaction comes through the top pad, 
it suffices to consider the couplings of lattice sites to this pad. The resulting susceptibility $\chi(\omega)\equiv q_0(\omega)/X(\omega)$ relating the perturbation $X$ to the charge $q_0$ at the top pad can be obtained by performing the Fourier transformation of the response function (see e.g., Ref.~\citenum{DPL92}):
 \eqn{
\chi(\omega,T)=\frac{\hbar g}{2e}\int_{-\infty}^{\infty}dt ~e^{i\omega t}
\frac{\langle{q}_{0}(t)\dot{{q}}_{0}(0)\rangle_{{\rm eq}}}{k_{\rm B} T},
}
where $\langle\cdots\rangle_{\rm eq}$ denotes the ensemble average over the steady-state regime.

Figure~3(b) in the main text shows numerically obtained ${\rm Im}\,\chi/2e$ as a function of temperature for the system with the $50 \times 50$ JJA and the parameters used in the experiment. The linewidths indeed show their peaks around the transition temperature $k_{B} T/E_{\rm J} \sim 0.91$ ($k_{B} T/E_{\rm J} \sim 0.43$) in the absence of fluxes (presence of half fluxes). 
Notably, it is expected that the peak temperatures of ${\rm Im}\,\chi/2e$ do not change significantly according to the damping rate $\Gamma$, whereas ${\rm Im}\,\chi/2e$ itself depends on $\Gamma$. 

Figure~\ref{FigS3} shows the size $L$ dependence of the BKT transition temperatures determined by the spatial correlation functions and the peak temperatures of the linewidths. For each $L$, the linewidth peak temperature coincides with the BKT transition temperature within their numerical errors, which indicates that the linewidth gives a dynamical signature of the BKT transition in the JJA.

\section*{Appendix C: Input-output relations}
\label{input_output}
Finally, to relate the susceptibility to an experimentally measured quantity, we invoke the input-output formalism. After employing the rotating-wave approximation ($X\to a/\sqrt{2}$), the equation of motion for the cavity field at  frequency $\omega$ can be obtained as \cite{cosmic2018circuit}
\begin{equation}
\begin{split}
-i\omega a&=-i\omega_c a-\frac{\kappa_{\rm ext}+\kappa_{\rm int}}{2}a+i\tilde{\chi} a
+\sqrt{\kappa_{\rm ext}} a_{\rm in},\\
\tilde{\chi}&=\frac{g}{2}\frac{\chi}{2e},
\end{split}
\end{equation}
where $\omega_c$ is the cavity frequency, 
$\kappa_{\rm ext}$ ($\kappa_{\rm int}$) is the external (internal) microwave field loss, and $a_{\rm in}$ is the input field. 
Using the input-output relation
\eqn{
\sqrt{\kappa_{\rm ext}}a=a_{\rm in}-a_{\rm out},
}
we obtain
\begin{equation}\label{seq:input-output}
\left\langle \frac{a_{\rm out}}{a_{\rm in}}\right\rangle =
1-\frac{\kappa_{\rm  ext}}{\frac{\kappa_{\rm ext}+\kappa_{\rm int}}{2}-i(\omega-\omega_c+\tilde{\chi})}.
\end{equation}
Decomposing the susceptibility into the real and imaginary parts as $\tilde{\chi}\equiv\tilde{\chi}_{r}+i\tilde{\chi}_{i}$, we arrive at the following equality relating the reflection coefficient (which has been experimentally measured) to the susceptibility $\chi$ (which can be obtained via the Langevin dynamic simulation):
\begin{equation}
\begin{split}
{\rm Re}\left[\left\langle \frac{a_{\rm out}}{a_{\rm in}}\right\rangle \right]&=
1-\frac{\kappa_{\rm ext}\kappa_{\rm tot}/2}{(\kappa_{\rm tot}/2)^2+(\omega-\omega_c+\tilde{\chi}_r)^2},\\
\kappa_{\rm tot}&=\kappa_{\rm ext}+\kappa_{\rm int}+2\tilde{\chi}_i.
\end{split}
\end{equation}
Since $\kappa_{\rm ext}$ and $\kappa_{\rm int}$ are almost temperature independent, it is evident that the temperature dependence of the resonance linewidth observed in the reflection coefficient originates from that of the imaginary part of the susceptibility.

\bibliography{main.bib}
\end{document}